\documentclass[runningheads]{llncs}
\usepackage[T1]{fontenc}
\usepackage{graphicx}
\begin{document}

\title{Deep Learning Ensemble for Predicting Diabetic Macular Edema Onset Using Ultra-Wide Field Color Fundus Image}
\author{Pengyao Qin\inst{1} \and Arun Thirunavukarasu\inst{2} \and Theodoros Arvanitis\inst{1} \and Le Zhang\inst{1}}
\authorrunning{Pengyao Qin et al.}
\institute{Digital Healthcare and Medical Imaging Research Group, \\School of Engineering, College of Engineering and Physical Sciences, \\University of Birmingham, Birmingham, UK 
\and Nuffield Department of Clinical Neurosciences, University of Oxford, Oxford, UK}
\date{}

\maketitle

\begin{abstract}
Diabetic macular edema (DME) is a severe complication of diabetes, characterized by thickening of the central portion of the retina due to accumulation of fluid. DME is a significant and common cause of visual impairment in diabetic patients. Center-involved DME (ci-DME) is the highest risk form of disease because fluid extends close to the fovea which is responsible for sharp central vision. Earlier diagnosis or prediction of ci-DME may improve treatment outcomes. Here, we propose an ensemble method to predict ci-DME onset within a year, after using synthetic ultra-wide field color fundus photography (UWF-CFP) images provided by the DIAMOND Challenge during development. We adopted a variety of baseline state-of-the-art classification networks including ResNet, DenseNet, EfficientNet, and VGG with the aim of enhancing model robustness. The best performing models were Densenet-121, Resnet-152 and EfficientNet-b7, and these were assembled into a definitive predictive model. The final ensemble model demonstrates a strong performance with an Area Under Curve (AUC) of 0.7017, an F1 score of 0.6512, and an Expected Calibration Error (ECE) of 0.2057 when deployed on the synthetic test dataset. Results from our ensemble model were superior/comparable to previous recorded results in highly curated settings using conventional fundus photography/ultra-wide field fundus photography. Optimal sensitivity in previous studies (using humans or computers to diagnose) ranges from 67.3\% $\sim$ 98\%, specificity from 47.8\% $\sim$ 80\%. Therefore, our method can be used safely and effectively in a range of settings may facilitate earlier diagnosis, better treatment decisions, and improved prognostication in ci-DME.


\end{abstract}

\section{Introduction}

Diabetic macular edema (DME) is a leading cause of vision loss among diabetic patients worldwide \cite{1}. It is characterized by the accumulation of fluid in the central portion of the retina (the macula), and can occur at any stage in the natural history of diabetic retinopathy (DR) \cite{2}. DME leads to severe vision impairment, although a variety of interventions are now available to prevent and even reverse disease progression if detected and treated early \cite{2}. Center-involved diabetic macular edema (ci-DME) is the highest-risk form of disease, as the central macula (the fovea) is responsible for central sharp vision. Clinically, DME is best characterised using optical coherene tomography (OCT), as high resolution cross-sectional images of the retina clearly reveal fluid accumulation and can be measured to assess response to treatment.

En-face fundus photography, 2D colour imaging of the retina, is a cheaper and more widely available modality than OCT, and is already used in screening programmes to detect diabetic retinopathy \cite{2}, \cite{varadarajan2020predicting}. However, clinicians cannot reliably diagnose DME on fundus photographs, and instead rely on surrogate markers of fluid accumulation such as hard exudate formation \cite{Heng2017}. These surrogates are poor predictors of DME, and screening could therefore miss patients that would benefit from early treatment, while simultaneously resulting in many unnecessary referrals to ophthalmologists for further investigation \cite{wangyut}. Computational analysis of fundus photographs can improve on clinical assessment, and thereby improve the utility of fundus photography to screen for DME.

Ultra-wide field color fundus photography (UWF-CFP) is a more advanced form of en-face fundus imaging that captures a wider portion of the retina. UWF-CFP may capture more features of DME than conventional photography, such as through higher resolution imaging of the macula or due to peripheral retinal consequences of mass effect exerted by fluid accumulation. The DIAMOND Challenge aimed to leverage this potential through development of artificial intelligence models capable of predicting the onset of ci-DME within one year based on individual UWF-CFP images. By focusing on predictive modeling, the challenge sought to shift the management paradigm from reactive to proactive treatment of ci-DME, thereby reducing the incidence of vision loss in diabetic retinopathy \cite{Gurung2023}. The DIAMOND Challenge also introduced an additional methodological complexity: only code (rather than model weights) was submitted, which the organizing committee then ran on a cloud-based cluster. This tasked participants with developing highly generalizable models with the necessary flexibility to perform in real-world situations where data heterogeneity, privacy, and logistical limitations are common.

In this work, we propose a deep learning ensemble-based approach to predict the development of clinically significant diabetic macular edema (ci-DME) using ultra-widefield color fundus photography (UWF-CFP) images. To achieve this, we employed several state-of-the-art convolutional neural networks (CNNs), including DenseNet, ResNet, EfficientNet, and VGG architectures, each of which offers unique strengths in feature extraction and classification. These models were trained with comprehensive data augmentation strategies to enhance robustness and reduce overfitting, particularly given the limitations of the dataset size and variability.

To further improve predictive performance, an ensemble method was introduced to combine predictions from the top-performing models. By leveraging the complementary strengths of individual networks, the ensemble approach aims to boost overall accuracy and reliability, offering a more generalizable solution for ci-DME prediction. This methodology addresses both classification and calibration challenges, ensuring that the predictions are not only accurate but also reliable for potential clinical applications. The ensemble strategy underscores the importance of combining diverse model architectures to achieve a more robust and clinically meaningful outcome.

\section{Method}
\subsection{Dataset}
The DIAMOND Challenge provided data sourced from 14 French hospitals as part of the EVIRED project, which aims to predict ci-DME development and anticipate the onset of DR complications. For evaluation, the DIAMOND Challenge incorporated independent datasets from Algeria alongside the French hospital data. This approach promoted universally applicable solutions, capable of serving diverse population groups and settings. During the coding period, a synthetic dataset generated using images from the Deep Diabetic Retinopathy Image Dataset (DeepDRiD) was provided to assist in the development and local testing of the algorithms. DeepDRiD is available under the Creative Commons Attribution Share Alike 4.0 International license, and served as a development and testing resource with ci-DME labels generated based on DR severity. This dataset includes 204 UWF-CFP in total, with 154 images partitioned for training and 50 images partitioned for validation. Also a one-patient dataset is provided, including real image data from one patient, with images taken by different UWF-CFP device (OPTOS and CLARUS) from each eye to enhance model development with actual ci-DME labels.

\subsection{Baseline Network}

We adopted four CNN architectures to develop a generalisable and robust model: Resnet, Densenet, EfficientNet and VGG.

\textbf{ResNets (Residual Networks)} are renowned for high performance in classification tasks, particularly in medical image analysis. A deep residual learning framework is employed to address the vanishing gradient problem and the degradation of network performance as depth increases. This approach enables the training of very deep networks by introducing residual mapping, which allows gradients to propagate through the network more effectively, enhancing training efficiency and performance \cite{he2016deep}.

\textbf{DenseNet (Densely Connected Convolutional Networks)} excels in efficient feature usage and parameter reduction. It introduces dense blocks where each layer connects to every other layer in a feed-forward manner, promoting feature reuse and mitigating the vanishing gradient problem. DenseNet is often preferred for its ability to achieve high performance with fewer parameters than with ResNets \cite{huang2017densely}.

\textbf{EfficientNet} scales network dimensions (depth, width, and resolution) uniformly using a compound scaling method. This model achieves superior performance on the ImageNet dataset and transfers well to other datasets, offering a balance of high accuracy and computational efficiency. EfficientNet's ability to maintain small model sizes while ensuring fast computation speeds makes it suitable for local applications such as hospital-based retinal image classification \cite{tan2019efficientnet}.

\textbf{VGG (Visual Geometry Group Network)} is an earlier convolutional neural network (CNN) architecture which uses small convolutional filters and deep networks (16-19 layers) to attain strong performance on large-scale image recognition tasks like ImageNet. VGG is known for its large model size and lengthy computation times relative to newer architectures, but remains a useful baseline model for image classification tasks such as object recognition and medical image classification \cite{simonyan15}

\subsection{Model ensemble}
Model ensembling combines the outputs produced by multiple models into a single prediction process, which can overcome shortcomings associated with individual estimators such as high variance, noise, and bias. In this work, three ensemble strategies were used. 

\textbf{Plurality Voting:} The first was plurality voting: where the class with the highest number of votes is used as the final prediction. All models contribute equally to the decision-making process, preventing dominance by any single model. Plurality voting treats all votes equally, ignoring the confidence levels of individual model \cite{evaluatingdeep}.

\textbf{Averaging:} The second was averaging: where the final outputted probability is the unweighted average of the probabilities estimated by each model. By averaging probabilities instead of class labels, this method incorporates the confidence of each model’s predictions, often leading to better-calibrated outputs. Averaging can produce smoother predictions, especially in cases of highly imbalanced data, as it reduces the influence of extreme or outlier predictions from individual models \cite{Ganaie_2022}.

\begin{figure}[t]
    \centering
    \includegraphics[width=\linewidth]{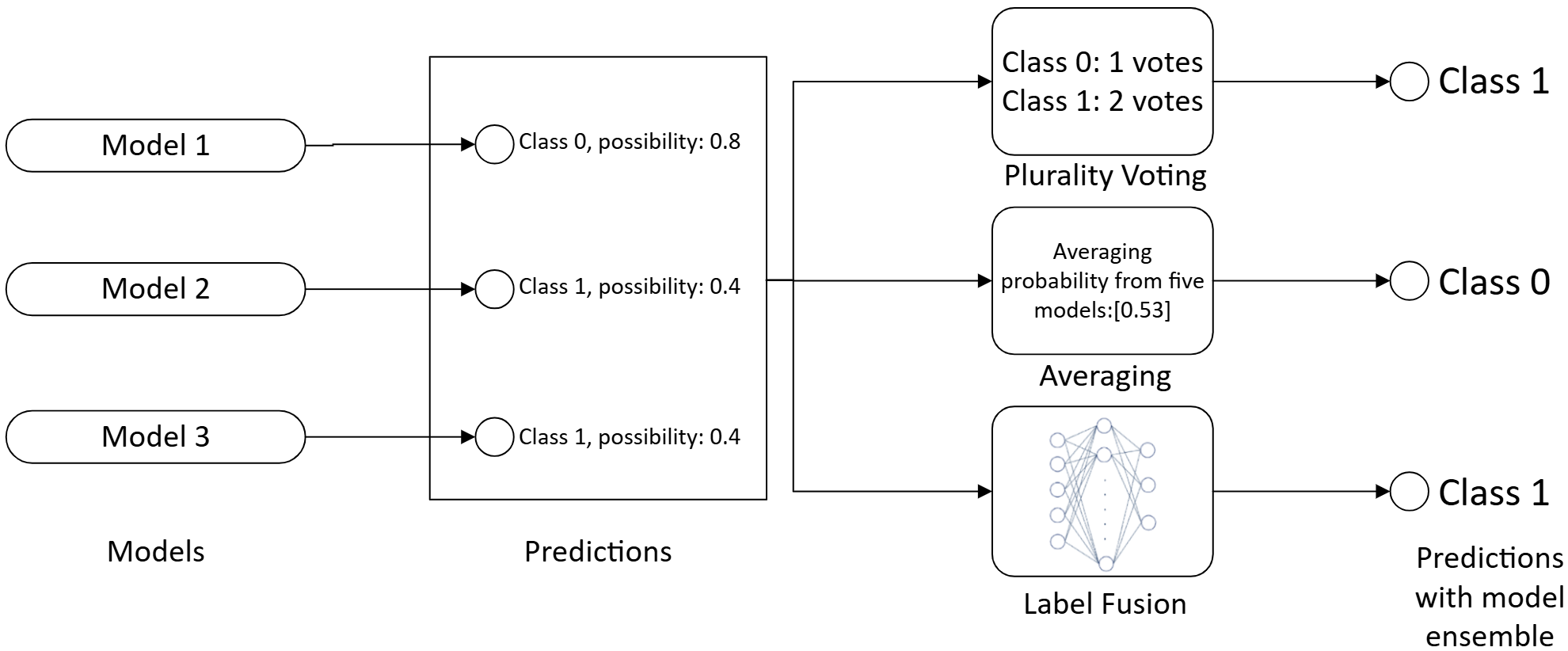}
    \caption{Diagram of an example of each ensemble strategy used in this study, with three individual models predicting ci-DME.} 
    \label{Fig1}
\end{figure}

\textbf{Label Fusion:} The third technique was label fusion using a three-layer simple neural network, which was trained using the predictions from several individual models as inputs, thereby learning to assign appropriate weights to each individual model. An example of these three strategies using five models and two classes is shown in Fig.~\ref{Fig1}. Final prediction outputs can vary widely with different ensemble strategies.

\section{Experiments}

\subsection{Experimental setting}

\begin{figure}[t]
    \centering
    \includegraphics[width=0.8\linewidth]{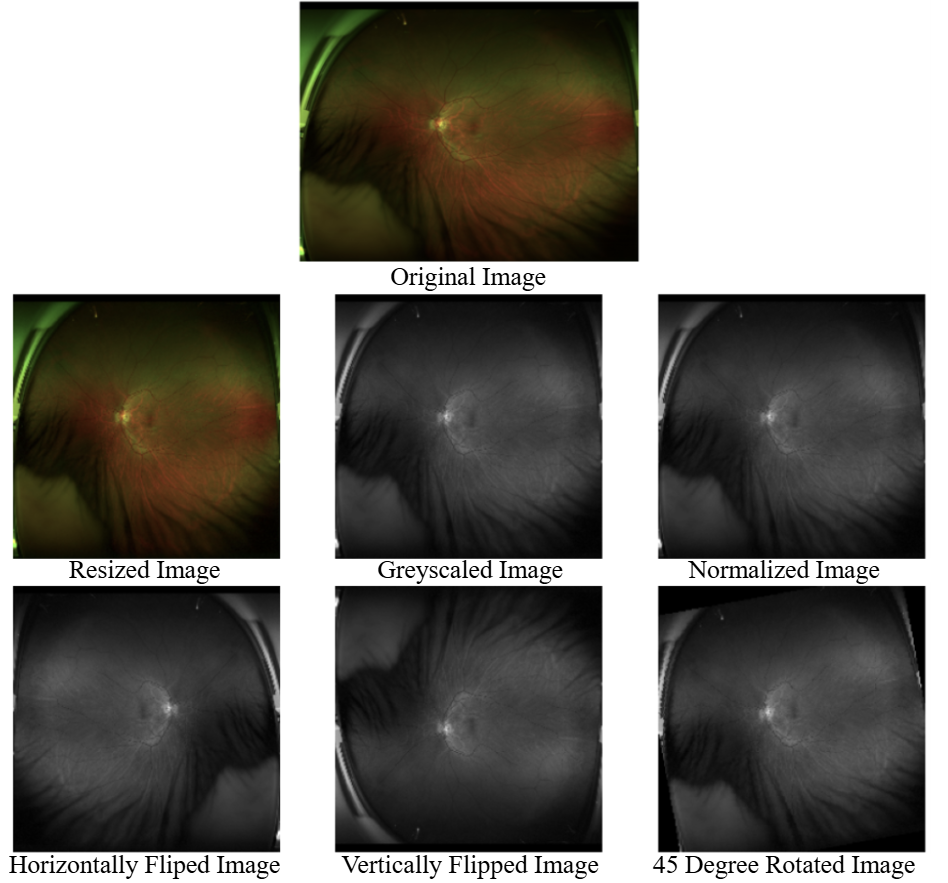}
    \caption{Examples of image processing employed during model development. Images were resized to 224 $\times$ 224 px, greyscaled, normalized, randomly reflected in the horizontal and vertical axis, and randomly rotated through 45\textdegree. The effects of each step are shown in each panel.}
    \label{Fig2}
\end{figure}

\textbf{Image Preprocessing.} 
In all experiments, the images were resized to 224 $\times$ 224 pixels from their original size. Other image augmentation techniques were also employed, including random horizontal reflection, random vertical reflection, and random rotation. Example images are shown in Fig.~\ref{Fig2}.

\textbf{Training Hyper-parameters.} Throughout the training process, the adam optimizer was used with a learning rate scheduler with exponential decay. Networks were trained using 200 epochs and the models with the best performance were saved. Performance was gauged with the metrics suggested by the DIAMOND Challenge: Area under the receiver operating characteristic curve (AUC), F1-score (F1) and Expected Calibration Error (ECE) \cite{estimatingexpectedcalibrationerrors}. 

\subsection{Evaluation Metrics}

For evaluation, we used the following metrics provided by the challenge organizer:

\textbf{Area under the curve (AUC):} The AUC is computed as the area under the receiver operating Characteristic (ROC) curve. AUC values range between 0 to 1; with 1 representing perfect classification performance across all model thresholds. The use of AUC aligns with the goal of achieving high sensitivity and specificity in predictions and was therefore weighted highest during evaluation and ranking. 

\textbf{F1 score and expected calibration error (ECE):} The F1 score and ECE were considered as secondary metrics (lower weighted) to further assess the relevance of computed probabilities. The F1 score is the harmonic mean of precision and recall, which is used to balance these two indicators. F1 helped ensures that binary predictions (when using a probability cutoff of 0.5) were relevant. F1 also assesses precision (or positive predictive value) which is not captured by AUC \cite{Grandini2020MetricsFM}. ECE measures the calibration of predicted probabilities, indicating the difference between predicted confidence and actual correctness. The calculation method of ECE is to divide the prediction distribution of the model output into multiple intervals, calculate the difference between the average prediction probability and the actual accuracy in each interval, and then sum them up to get the ECE value, lower ECE values indicate better calibration of the model. The use of ECE helped calibrate model properties for the context in which it will be used, such as by ensuring that its predicted probabilities matched the actual probabilities of the ground truth distribution \cite{estimatingexpectedcalibrationerrors}. The following performance function was used to determine which model was saved.
\[ S = AUC + 0.5 \cdot F1 + 0.5 \cdot (1-ECE) \]

\subsection{Model evaluation}

The results in Table.~\ref{tab1} compare the performance of 5 baseline models, with the DenseNet-121 achieving the highest overall score (1.5262) due to its strong AUC (0.7617), F1 score (0.7347), and low ECE (0.2057), reflecting its efficiency in classification and calibration. ResNet-152 also performed well, with the highest AUC (0.7633) and a solid overall score (1.47875), although its lower F1 score (0.6275) limited its effectiveness. EfficientNet-b7 demonstrated a balance between metrics, with a respectable overall score of 1.46325, but was outperformed by DenseNet-121. ResNet-50 and VGG-19, while adequate, lagged behind in overall scores (1.4449 and 1.4231, respectively), showing limitations in calibration and predictive accuracy compared to the more modern architectures.

Table.~\ref{tab2} shows the performances of the ensemble methods, evaluated on the internal testing dataset. For the individual evaluation, label fusion obtained the best performances (overall score of 1.4343) comparing to the other models ensemble methods. The results in Table.~\ref{tab2} highlight the strengths and weaknesses of three ensemble methods for ci-DME prediction. Plurality voting, while simple and well-calibrated (ECE = 0.1023), achieved the lowest AUC (0.6517), F1 score (0.5714), and overall score (1.3862), indicating limited predictive capability. Averaging improved AUC (0.71885) and F1 score (0.6862) significantly, showcasing better classification performance, but it suffered from poor calibration (ECE = 0.5361), which lowered its overall score to 1.2939. Label fusion outperformed the other methods with the highest overall score (1.4343), offering a balanced performance with competitive AUC (0.7017), F1 score (0.6512), and improved calibration (ECE = 0.2860), demonstrating its ability to effectively integrate model outputs.

\setlength{\tabcolsep}{4mm}
\renewcommand{\arraystretch}{1.1}
\begin{table}[t]
\caption{Experimental results for each baseline model, with inidividual evaluation metrics presented alongside overall performance scores. Densenet-121 exhibited the highest performance score.}\label{tab1}
\centering
\begin{tabular}{|l|l|l|l|l|}
\hline
Network & AUC & F1 Score & ECE & Overall Score \\ \hline
Resnet-50 \cite{he2016deep} & 0.7233 & 0.7778 & 0.3346 & 1.4449\\ \hline
Resnet-152 \cite{he2016deep} & 0.7633 & 0.6275 & 0.1966 & 1.47875 \\ \hline
Densenet-121 \cite{huang2017densely} & 0.7617 & 0.7347 & 0.2057 & 1.5262 \\ \hline
EfficientNet-b7 \cite{tan2019efficientnet} & 0.7467 & 0.6780 & 0.2449 & 1.46325 \\ \hline
VGG-19 \cite{simonyan15} & 0.7350 & 0.6154 & 0.3392  & 1.4231  \\   \hline
\end{tabular}
\end{table}


\setlength{\tabcolsep}{4mm}
\renewcommand{\arraystretch}{1.1}
\begin{table}[t]
\caption{Experiment results about the evaluation metrics and overall scores of each model ensemble methods.}\label{tab2}
\centering
\begin{tabular}{|l|l|l|l|l|}
\hline
Ensemble method & AUC & F1 Score & ECE & Overall Score \\ \hline
Plurality voting & 0.6517 & 0.5714 & 0.1023 & 1.3862 \\ \hline
Averaging & 0.71885 & 0.6862 & 0.5361 & 1.2939 \\ \hline
Label Fusion & 0.7017 & 0.6512 & 0.2860 & 1.4343 \\ \hline
\end{tabular}
\end{table}


\section{Discussion and Conclusion}

In this study, we developed an ensemble-based deep learning model to predict the onset of ci-DME within 12 months using UWF-CFP images. Our model combined several CNN architectures and exhibited performance comparable to previous studies involving human experts and individual deep learning models diagnosing existing DME on fundus photographs \cite{wangyut}, \cite{1}. Our ensemble method ensured that classification decisions were not limited by any single model's weaknesses. This helped reduce the incidence of false negative classification, which in the real world could lead to patients missing out on crucial early treatment; as well as false positive classification, which can lead to overmedicalisation and unnecessary treatment. While the ensemble model’s overall performance was slightly lower than that of the best individual model (Densenet-121), it exhibited improved generalizability across diverse datasets. This trade-off--sacrificing peak performance for better generalization--makes the ensemble model a more robust application for clinical deployment. Though higher performance has been achieved for detecting existing ci-DME from fundus photographs (rather than OCT, the clinical standard for diagnosis) \cite{varadarajan2020predicting}, a predictive model like ours could help guide preemptive decision-making regarding active surveillance and early intervention. In the future, emerging attention-based models such as vision transformers (ViTs) could be integrated into the ensemble \cite{article}. These models have demonstrated superior performance in image classification tasks driven by selective focus on relevant portions of the image, which is highly relevant to medical tasks like ci-DME prediction where pathological changes are localised.

\begin{figure}[t]
    \centering
    \includegraphics[width=0.7\textwidth]{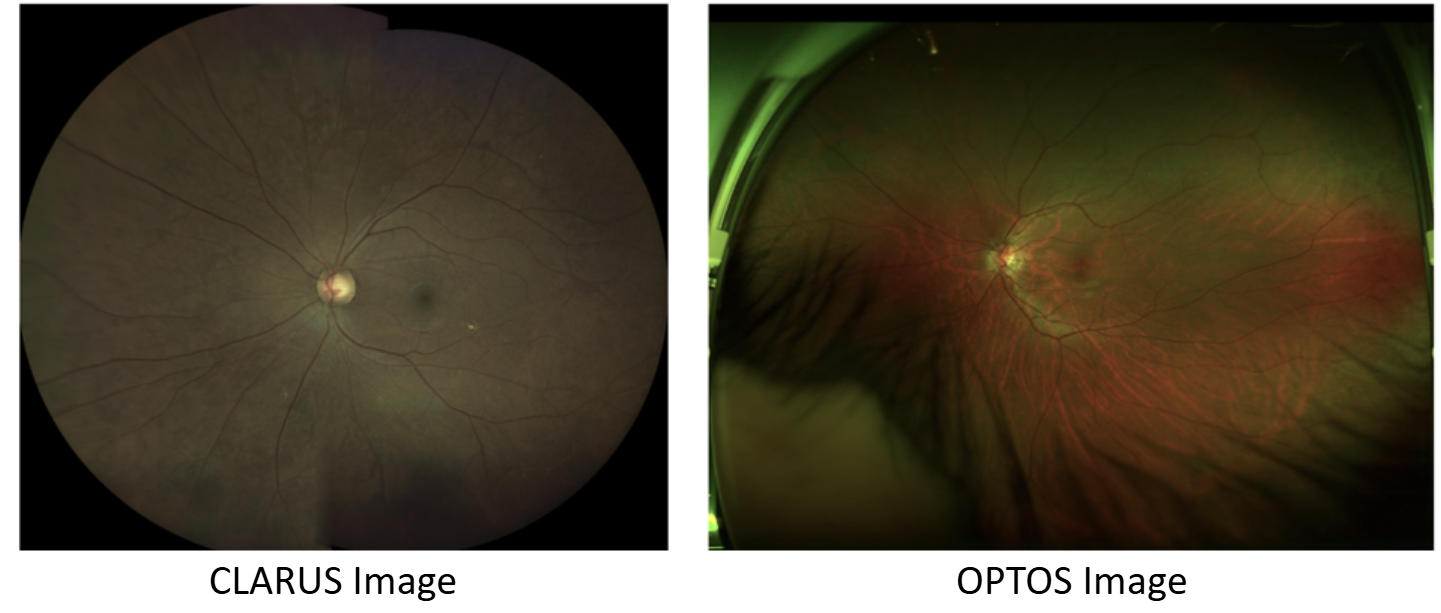}
    \caption{Examples of Ultra Wide Field Colour Fundus Photographs captured by different devices, Zeiss CLARUS (left), and OPTOS (right). The Challenge uses datasets from these two devices for evaluation. Significant differences in dimensions, resolution, and artifact can confound classification and thereby place higher demands on the generalisability of a model that is planned to be trained and/or tested across both modalities.}
    \label{Fig3}
\end{figure}

In the DIAMOND Challenge, synthetic datasets were used exclusively for the initial training phase in which baseline networks were established. While this approach offers certain advantages, such as controlled data characteristics and reduced ethical concerns (particularly regarding data privacy), it introduces critical limitations that may limit model performance on real-world data. During model ensembling, certain strategies demonstrated poor results, which we hypothesize may be attributable to the characteristics of the synthetic data. For example: synthetic datasets may overemphasize specific features or lack challenging edge cases, leading to models that do not complement each other well in an ensemble. The absence of realistic noise in synthetic datasets can cause ensemble methods that rely on diverse model outputs to underperform, as the models may converge on similar, biased predictions \cite{Hao2024SyntheticDI}. To mitigate these limitations, future iterations of the study should incorporate real-world datasets in the training and validation stages.

Timely intervention is associated with substantial improvement in patient outcomes, but fundus photography is insufficient to replace OCT in diagnosing ci-DME and regular OCT screening is associated with prohibitive costs and inaccessibility, particularly in lower income countries \cite{2}, \cite{varadarajan2020predicting}. Predictive deep learning models with UWF-CFP images may represent a cost-effective and scalable alternative, allowing more patients around the world to receive critical treatment as required \cite{vit}. This potential aligns well with the aims of the DIAMOND Challenge, which emphasized the value in generalisable models with potential to augment real-world clinical practice. Our approach aimed to maintain sufficient flexibility to cope with training and deployment across conventional fundus photography as well as UWF-CFP, with differences in frame of view, image artifact, and resolution that can confound classification (as shown in Fig.~\ref{Fig3}). In the 7-day period provided by the DIAMOND Challenge, we implemented a streamlined training and evaluation workflow using a bash script. This script automated the training of each component network separately on the available data. By using an ensemble method combined with model selection based on objective evaluation metrics, we ensured that the final submission was both robust and generalizable, potentially performing well even when applied to datasets derived from other imaging devices.

In conclusion, our ensemble model demonstrates the potential to significantly advance the early detection of ci-DME through the use of synthetic data before local deployment with authentic data from patients. Improved predictive models could facilitate proactive management and prompt treatment to improve outcomes in patients with or at risk of ci-DME. By integrating deep learning models with UWF-CFP imaging, accurate and accessible screening could promote timely intervention and ultimately better clinical outcomes for patients with DR.

\bibliographystyle{unsrt}
\bibliography{references}

\end{document}